# Digitization of 93.3-keV recoilless γ-radiation

Y. V. Radeonychev[1,2,*] and I. R. Khairulin[1,2]

[1] *Gaponov-Grekhov Institute of Applied Physics of the Russian Academy of Sciences, 46 Ulyanov Street, Nizhny Novgorod, 603950, Russia*

[2] *Lobachevsky State University of Nizhny Novgorod, 23 Gagarin Avenue, Nizhny Novgorod, 603950, Russia*

We propose a technique for transforming the intensity of quasi-monochromatic recoilless radiation with a photon energy of 93.3 keV, emitted by radioactive Mössbauer sources $^{67}$Ga or $^{67}$Cu, into a sequence of short pulses with individually and independently controlled, on demand, the moments of appearance of pulses, as well as the peak intensity, duration and shape of each pulse. The technique is based on the transmission of recoilless (Mössbauer) 93.3-keV photons from the source through a medium containing resonantly absorbing $^{67}$Zn nuclei. The pulses are formed due to the rapid reciprocating movement of the source relative to the absorber at specified moments in time along the direction of photon propagation at a distance not exceeding half the radiation wavelength. The produced sequences of γ-ray pulses are similar to the digitization of information carried by electromagnetic waves. They can also be used to develop Mössbauer spectroscopy of the atomic and subatomic structures, as well as open novel opportunities for γ-ray quantum optics.

*corresponding author: radion@appl.sci-nnov.ru

## I. INTRODUCTION

Methods for generating coherent electromagnetic radiation in the form of short pulses with controlled spectral and temporal characteristics in various frequency ranges from radio waves to X-rays are in high demand and are being intensively developed due to numerous applications in science and technologies. Radiofrequency and optical radiation in the form of pulse sequences with controlled characteristics is typical for digital communication networks and is the basis for the transmission of quantum information. Pulsed radiation of the X-ray and γ-ray domain with controlled parameters is attractive for applications in quantum communications and processing due to the potential for focusing to an angstrom scale, long-lived nuclear resonances for "photon storage" in nuclear coherence, as well as efficient detection of high-energy photons.

Presently, there are lots of methods for producing laser pulses with the required characteristics. However, it is much more difficult to control X-ray radiation with the photon energy of tens of keV due to mainly the short wavelength of such radiation. Indeed, laser radiation effectively interacts with atomic electrons whose optical properties are changed on demand by various techniques. On the contrary, the main effects of interaction of angstrom-wavelength photons with atoms are their ionization and recoil. As a result, most approaches well suited for controlling laser light are not feasible for hard X-rays.

At the same time, high-energy photons can get into resonance with quantum transitions of atomic nuclei and efficiently interact with them without recoil (Mössbauer interaction) similar to the interaction of optical photons with atomic electrons. In addition, unlike atomic transitions, recoilless nuclear resonances are usually almost naturally broadened even at room temperature, and their spectral width can be very small. For example, the 14.4-keV recoilless nuclear transitions in $^{57}$Fe nuclide normally have several megahertz linewidths at room temperature corresponding to enormous Q-factor on the order of $10^{12}$ [1]. Two orders of magnitude narrower linewidth (several dozen kilohertz) have 93.3-keV recoilless transitions in $^{67}$Zn nuclei corresponding to Q-factor on the order of $10^{15}$ [1-7]. This, in particular, serves as the basis for extremely precise spectroscopic measurements of the nucleus quantum transition frequencies, which are very sensitive to their microscopic environment and, therefore, can reveal the atomic and subatomic structure of the substance in which they are embedded.

High frequencies of the keV nuclear transitions make the Doppler effect a very efficient tool for measuring and controlling the frequencies of these transitions. For example, the movement of $^{67}$Zn absorber relative to the source at a constant velocity of only 170 nm/s along the 93.3-keV photon propagation direction shifts the position of its 13 kHz-wide spectral line by the linewidth, taking the absorber out of resonance with the source and dramatically changing the absorber opacity. These properties of nuclei underlie the uniquely high sensitivity and precision of Mössbauer spectroscopy in the study of matter [4,7]. They also give rise to novel mechanisms and methods for acoustic control of X-ray and γ-ray photons via resonant interaction with atomic nuclei [5,6], as well as open up prospects for the development of a flexible interface between X-ray photons and nuclei.

A relatively long decay time of the recoilless radiating and absorbing keV nuclear transitions makes it possible to significantly change both the spectrum and the time dependence of intensity of the emitted radiation by means of transmitting photons through a resonant absorber that rapidly moves relative to the source (or vice versa) at certain times or periodically, much faster than the radiative decay time of the corresponding nuclear transitions ([5,6,8-16] and references therein). This is because the motion causes, due to the Doppler effect, a temporal phase shift between the coherent nuclear response (coherent forward-scattered field) produced by the absorbing nuclei, and the incident field. This phase shift changes the character of the interference between the incident and forward-scattered fields [5,6,8-16], which underlies the resonant interaction between radiation and nuclei.

Such an acoustic control of γ-radiation intensity was first implemented for 93.3-keV photons [5,6]. It was shown that the stepwise displacements of the $^{67}$Ga source by a distance, multiple of a half the photon wavelength, resulted in transformation of the γ-radiation intensity into short pulses in the motionless optically thick resonant $^{67}$Zn absorber [5,6]. The intensity of the transmitted field achieved its maximum at the moments when the distance between the source and absorber changed by $(2n+1)\lambda_s/2$ (where $\lambda_s$ is the photon wavelength and $n \in \mathbb{N}$) and a minimum at the moments when the distance between the source and absorber changed by $n\lambda_s$. At these moments, constructive or destructive interference, respectively, of incident and coherently scattered fields took place [5,6].

Similar pulses were obtained in the case of 14.4-keV radiation emitted by $^{57}$Co source [8-12], when the source [8-10] or the $^{57}$Fe absorber [11,12] was stepwise displaced at certain moments in time. It is also worth mentioning the production of a regular sequence of short γ-ray pulses by means of a rapid periodic displacement (vibration) of the absorber relative to the source with a frequency significantly exceeding the decay rates of the corresponding nuclear transitions [13-15]. In this case, the forward-scattered field becomes phase-modulated [15]. Therefore, properly adjusted frequency and amplitude of modulation (determined by the frequency and amplitude of the absorber vibration), as well as the detuning between the frequencies of the source and absorber result in different pulse sequences [13,14], as well as can lead to suppression of resonant absorption [16].

There are two commercially available types of the 93.3-keV radiation sources, namely the sources based on the $^{67}$Ga and $^{67}$Cu radionuclides (Fig. 1). There are a number of techniques for obtaining different compounds with $^{67}$Ga or $^{67}$Cu that produce 93.3-keV radiation of various intensity and spectral content [1-6,17-20]. Earlier results with $^{67}$Zn were obtained mostly using $^{67}$Ga sources [2-7]. The recent breakthrough in the production of $^{67}$Cu with high specific activity for the needs of radioimmunotherapy and single-photon emission computed tomography in medicine [17,18] has opened up novel opportunities for the Mössbauer γ-ray optics with $^{67}$Zn.

In this paper, using the example of the $^{67}$Ga or modern $^{67}$Cu radioactive source and the $^{67}$Zn absorber, we propose a technique which allows one to independently control the temporal characteristics of γ-radiation emitted by the source. Unlike earlier studies [5,6,9-12], this technique is based on a rapid *reciprocating* displacement of the source relative to the resonant absorber (or vice versa) along the direction of photon propagation at a distance less than or equal to half the radiation wavelength. We show that this type of motion makes it possible to (i) transform a constant intensity of 93.3-keV radiation into variable number of pulses including a single pulse; (ii) individually adjust the peak intensity of each pulse; (iii) individually adjust the duration of each pulse; (iv) adjust the individual time interval between the adjacent pulses in a sequence; (v) produce on-demand an individual shape of each pulse.

The paper is organized as follows. In Section II, we describe the model and derive a solution for the intensity of radiation transmitted through an optically deep recoilless resonant $^{67}$Zn absorber that can move relative to the source (or vice versa) along the direction of photon propagation. In Section III, we briefly describe the physical picture of the resonant absorption of the single-photon field in the optically thick resonant absorber as an interference between the incident and forward-scattered fields. In Section IV using the example of a rapid reciprocating displacement of the source relative to the absorber, we show that this interference underlies the transformation of the field in the resonant absorber. We derive an analytical solution for the transmitted intensity in the case of sufficiently rapid reciprocating displacement of the source and show how the characteristics of the displacement are mapped to the characteristics of the produced pulses. In Section V we summarize the results.

## II. THEORETICAL MODEL

Let us consider the following experimental conditions similar to those implemented in [5,6]. Radiation with a photon energy of 93.3 keV is emitted by either $^{67}$Ga or $^{67}$Cu radioactive Mössbauer source (Fig. 1, left side). Both types of the source are made in the form of micron-thickness foil mounted on a piezoelectric transducer. At certain moments in time, the pulsed voltage supplied to the transducer causes a reciprocating movement of the source relative to the absorber. Alternatively, the absorber can be moved relative to the source with the same result [11-16]. The radioactive source stochastically emits 93.3-keV Mössbauer single photons separated in time (Fig. 1, left side). The photons propagate through a of recoilless resonant $^{67}$Zn nuclei (Fig. 1, right side) and are detected behind the absorber. The moments of photon detection (collected in the corresponding time bins) are measured by a laboratory clock. The measurement lasts for a fixed time interval, which starts from the moment the clock is turned on and ends when the clock is turned off. This procedure is repeated many times and gives the time dependence of the number of photons leaving the absorber per unit of time, which is proportional to the time dependence of intensity of the transmitted radiation.

Calculation of the intensity of electromagnetic field $I^{(av)}(t)$ measured by this technique is convenient to begin from the intensity associated with the detection of a single photon, $I(t,t_0)$. The intensity $I(t,t_0)$, is proportional to the time dependence of the probability for detecting the 93.3-keV photon per unit of time starting from the moment $t_0$ when the source nucleus appears in the radiating state $|b\rangle$, opening the possibility for the 93.3-keV photon to be emitted (Fig. 1) [9-16]. The intensity associated with the single photon emitted by the excited $|b\rangle$ of the source can be written in the form [5,6, 9-16]

$$I_s(t,t_0) \propto \theta(t-t_0)\exp\left[-(t-t_0)\Gamma_s\right], \qquad (1)$$

where $\theta(\tau)$ is the unit step function and $\Gamma_s$ is the decay rate of the source state $|b\rangle$.

In the experimental conditions considered here, the random moment $t_0$ cannot be detected and is averaged,

$$I^{(av)}(t) = \int_{-\infty}^{t} N(t_0) I(t,t_0) dt_0, \qquad (2)$$

where $N(t_0)$ is the average number of photons emitted by the radioactive source per unit of time in the direction of detection. Similar to [5,6], it is assumed below that $N(t_0) = N = Const$.

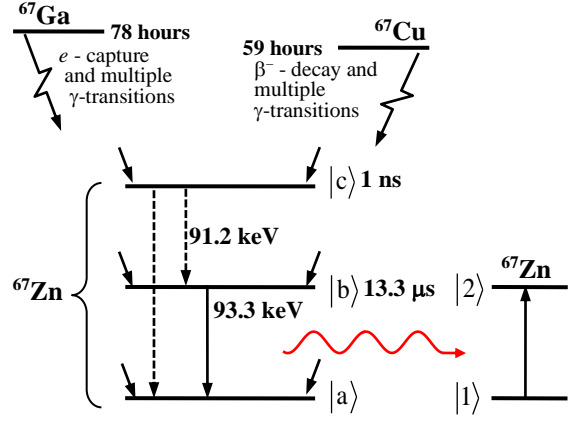

**Fig. 1.** Scheme of the radioactive decay of $^{67}$Ga and $^{67}$Cu sources (left side) and the resonant absorption of 93.3-keV photons by $^{67}$Zn nucleus of the absorber (right side). The $^{67}$Ga nucleus decays (inclined arrows) via electron capture, transforming into the $^{67}$Zn nucleus in five energy states, including ground state [21]. The $^{67}$Cu nucleus undergoes the $\beta^-$ decay, transforming into the $^{67}$Zn nucleus in four energy states (inclined arrows), including ground state [19,20]. In our case, only the ground state $|a\rangle$ and two excited states $|b\rangle$ and $|c\rangle$ are important. Excited states of $^{67}$Zn nucleus then radiatively decay to the lower energy states (vertical arrows). For both $^{67}$Ga and $^{67}$Cu, the probability to decay into the state $|b\rangle$ and state $|c\rangle$ is of the same order [19-21]. Therefore, an essential part of the 93.3-keV photons emitted at the transition $|b\rangle \rightarrow |a\rangle$ does not correlate with the 91.2-keV photons, corresponding to the transition $|c\rangle \rightarrow |b\rangle$

According to [5,6, 9-16,22,23] at the entrance to the $^{67}$Zn absorber, the electric field of the 93-keV single-photon wave packet, corresponding to the intensity (1), can be represented in the form of classical wave:

$$E_s(t,t_0) = E_0 \theta(t-t_0) e^{-(i\omega_s + \gamma_s)(t-t_0) + i\varphi_0}, \qquad (3)$$

where $E_0$ is the field amplitude, $\gamma_s = \Gamma_s/2$ is the half-width of the source spectral contour, corresponding to the lifetime of the state $|b\rangle$, $1/\Gamma_s \approx 13.3$ μs, $\omega_s$ is the carrier frequency of the field corresponding to the wavelength $\lambda_s \approx 0.13$ Å, and $\varphi_0$ is the random initial phase of the field.

The 93-keV single-photon wave packet propagates through a resonant Mössbauer absorber $^{67}$Zn (Fig. 1, right part). The $^{67}$Zn nuclei can be embedded into metal Zn or different monocrystalline (foil) or polycrystalline (powder) compounds such as ZnTe, ZnSe, ZnS, ZnO, ZnF$_2$ [4,7]. All absorbers can be utilized within the same experimental setup, in which the absorber is fixed (or moves at a constant velocity to tune into resonance with the source) and the source is mounted to a Mössbauer piezoelectric transducer [5,6]. In this case the source can move as a whole (within the beam cross-section, where photons interact with the absorber nuclei) along the photon propagation direction with a displacement function $S_{shift}(t)$ relative to the absorber.

Due to the source movement, the single-photon field incident to the absorber becomes phase-modulated because of the Doppler effect:

$$E_{in}(t,t_0) = E_s(t,t_0) e^{ik_s S_{shift}(t)}, \qquad (4)$$

where $k_s = 2\pi/\lambda_s$. Then the photon field at the exit from the resonant absorber with the Mössbauer (optical) thickness $T_a$ and the half-width of the absorption line $\gamma_a$ can be calculated as the convolution integral of the incident field (4) and the absorber response function $a(t)$ in the form [5,6,8-12]:

$$E_{out}(t,t_0) = \int_{-\infty}^{\infty} a(t-\tau) E_{in}(\tau,t_0) d\tau, \qquad (5)$$

with

$$a(t) = e^{-T_e/2} \left[ \delta(t) - T_a \gamma_a e^{-(i\omega_a+\gamma_a)t} \theta(t) J_1\left(\sqrt{2T_a\gamma_a t}\right) / \sqrt{2T_a\gamma_a t} \right]. \qquad (6)$$

In (6) $\delta(t)$ is the Dirac delta function, $J_1(x)$ is the Bessel function of first kind of the first order, $\omega_a$ is the frequency of the resonant transition $|1\rangle \leftrightarrow |2\rangle$ of the $^{67}$Zn absorber (Fig. 1, right side), which can differ from the central frequency of the source, $\omega_s$, due to isomer shift or Doppler shift produced by a motion of the source relative to the absorber with a constant velocity, and $T_e$ is the exponent factor of the non-resonant attenuation associated with photoelectric absorption and incoherent scattering. In what follows, we assume for simplicity that the photon field (4) is tuned in resonance with the quantum transition $|1\rangle \leftrightarrow |2\rangle$ of the absorber, $\omega_s = \omega_a \equiv \omega$, and the widths of the spectral contours of the source and absorber are the same, $\gamma_s = \gamma_a \equiv \gamma$. Then Eq. (5) can be rewritten as:

$$E_{out}(t,t_0) = E_s(t,t_0) e^{-T_e/2} \left[ e^{ik_s S_{shift}(t)} + A_{AR}(t,t_0) \right], \qquad (7)$$

where

$$A_{AR}(t,t_0) = -T_a \gamma \int_{t_0}^{t} \frac{J_1\left(\sqrt{2T_a\gamma(t-\tau)}\right)}{\sqrt{2T_a\gamma(t-\tau)}} e^{ik_s S_{shift}(\tau)} d\tau. \qquad (8)$$

According to (7) (except for the term $e^{-T_e/2}$), the single-photon field at the exit from the absorber is the sum of the incident field (4) (the first term in (7)), and the absorber response often called the coherent forward-scattered field by the absorber nuclei (the second term in (7)).

Using (7), the intensity of the single-photon pulse at the exit from the absorber, $I_{out}(t,t_0) = c|E_{out}(t,t_0)|^2/(8\pi)$, can be written in the form

$$I_{out}(t,t_0) = I_s(t,t_0) e^{-T_e} \left\{ \left(1 - |A_{AR}(t,t_0)|\right)^2 + 4|A_{AR}(t,t_0)| \cos^2\left[0.5 k_s S_{shift}(t) - 0.5 \arg(A_{AR}(t,t_0))\right] \right\}, \qquad (9)$$

where, according to (1), $I_s(t,t_0) = (cE_0^2/(8\pi))\theta(t-t_0)\exp[-2\gamma(t-t_0)]$ is the source radiation intensity in the direction of the absorber and $A_{AR} = |A_{AR}|\exp[i\arg(A_{AR})]$. Thus, according to (2), the measured intensity of the 93.3-keV radiation as a whole at the exit from the absorber is calculated by averaging intensity (9) of the single-photon field over $t_0$,

$$I_{out}^{(av)}(t) = N\int_{-\infty}^{t} I_{out}(t,t_0)dt_0. \qquad (10)$$

### III. ABSORPTION AS AN INTERFERENCE BETWEEN THE INCIDENT AND FORWARD-SCATTERED FIELDS

As follows from (7) and (8), for the propagating single-photon wave packet, the forward-scattered field at the beginning after its appearance is antiphase with respect to the incident field and increases gradually. This is most clearly seen when the source is at rest, $S_{shift}(t) \equiv 0$. In this case the amplitude of the forward-scattered field (8) takes the form

$$A_{AR}^{(rest)}(t,t_0) = \left[1 - J_0\left(\sqrt{2T_a\gamma(t-t_0)}\right)\right]e^{i\pi}, \qquad (11)$$

where $t \geq t_0$ is assumed.

As follows from (11), in general, the forward-scattered field oscillatory depends on time. Starting at $t = t_0$, it gradually increases with the characteristic rate $1/\tau_a$, where $\tau_a$ is the interval during which the Bessel function $J_0(x)$ in (11) changes from unit at $t = t_0$ to zero at $t - t_0 = \tau_a \approx 2.9/(\gamma T_a)$. As follows from (11) and (7), at $t - t_0 = \tau_a$ one has $J_0(\tau_a) = 0$, and the amplitude of the forward-scattered field becomes equal to the amplitude of the incident field. This results in disappearance of the output field. In the optically thick absorber, the attenuation rate $1/\tau_a$ of the single-photon field (7), (9) can significantly exceed the decay rate $\gamma_s = \gamma$ of the incident single-photon field (3), which is called the speed-up effect [24]. The oscillating amplitude of the forward-scattered field (11) leads to the oscillatory attenuation of the transmitted single-photon field in time called dynamical beats [24].

For the photon stream as a whole, integration (10) in the case of the motionless source averages the coherent forward-scattered single-photon field (11) and blurs the dynamical beats. As a result, the average output intensity is constant and less than the source intensity, $I_{out,rest}^{(av)}(t) < I_s^{(av)} = cE_0^2 N/(16\pi\gamma)$. However, if the source moves relative to the absorber, $S_{shift}(t) \neq 0$, the time dependence of the absorber forward-scattered field (8) causes the time dependence in the output field intensity (10).

### IV. PULSE FORMATION VIA RAPID RECIPROCATING DISPLACEMENT OF THE SOURCE

Let us consider the case of a rapid reciprocating motion of the source, namely, when the initially motionless source at a certain moment starts to move forward and backward along the field propagation direction, so that in a time interval $\Delta t_{total}$ it returns to the original position and stops. This reciprocating displacement of the source occurs so fast that $\Delta t_{total} \ll \tau_a$. Then, as follows from (8), the absorber response does not have time to change after the incident field and hence can be approximated by the relation $A_{AR}(t,t_0) = A_{AR}^{(rest)}(t,t_0)$, where $A_{AR}^{(rest)}(t,t_0)$ is described by equation (11). In this case, substitution of (9) and (11) into (10) gives the transmitted intensity in the form

$$I_{out}^{(av)}(t) = I_s^{(av)} e^{-T_e} \left[ e^{-T_a/2} I_0(T_a/2) + 4\left(1 - e^{-T_a/4}\right) \sin^2\left(\pi S_{shift}(t)/\lambda_s\right) \right], \quad (12)$$

where $I_0(x)$ is the modified Bessel function of the zeroth order. As follows from (12), the intensity of the transmitted photon beam maps the rapid reciprocating displacement function $S_{shift}(t)$ of the source with mapping function $\sin^2\left[\pi S_{shift}(t)/\lambda_s\right]$. This makes it possible to transform radiation from the source of constant intensity into various pulse sequences, including single pulse, with controlled amplitudes, durations and shapes. To show this, let us consider a model displacement function of the source, in the form of a sequence of short piecewise linear functions (Fig. 2a),

$$S_{shift}(t) = \sum_{i=1}^{M} S_{shift}^{(i)}(t). \quad (13)$$

Namely, each displacement $S_{shift}^{(i)}(t)$ starts at the moment $t_{start}^{(i)}$, then during the interval $\Delta t_f^{(i)}$ the source is shifted relative to the absorber by the amplitude $\Delta z^{(i)}$ in forward direction, then during the interval $\Delta t_c^{(i)}$ the source is at rest relative to the absorber, and then returned to its original position during the interval $\Delta t_b^{(i)}$:

$$S_{shift}^{(i)}(t) = \begin{cases} 0, & t < t_{start}^{(i)}, \\ \dfrac{\Delta z^{(i)}}{\Delta t_f^{(i)}}(t - t_{start}^{(i)}), & t_{start}^{(i)} \leq t < t_{start}^{(i)} + \Delta t_f^{(i)}, \\ \Delta z^{(i)}, & t_{start}^{(i)} + \Delta t_f^{(i)} \leq t < t_{start}^{(i)} + \Delta t_f^{(i)} + \Delta t_c^{(i)}, \\ \Delta z^{(i)} - \dfrac{\Delta z^{(i)}}{\Delta t_b^{(i)}}(t - t_{start}^{(i)} - \Delta t_f^{(i)} - \Delta t_c^{(i)}), & t_{start}^{(i)} + \Delta t_f^{(i)} + \Delta t_c^{(i)} \leq t < t_{start}^{(i)} + \Delta t_f^{(i)} + \Delta t_c^{(i)} + \Delta t_b^{(i)}, \\ 0, & t \geq t_{start}^{(i)} + \Delta t_f^{(i)} + \Delta t_c^{(i)} + \Delta t_b^{(i)}. \end{cases} \quad (14)$$

As follows from (12)-(14), in this case, the $i$-th pulse begins at $t_{start}^{(i)}$ and ends at $t_{start}^{(i)} + \Delta t_{total}^{(i)}$, where $\Delta t_{total}^{(i)} = \Delta t_f^{(i)} + \Delta t_c^{(i)} + \Delta t_b^{(i)}$ is the total duration of the $i$-th displacement. It has amplitude

$$I_{peak}^{(i)} = I_s^{(av)} e^{-T_e} \left[ e^{-T_a/2} I_0(T_a/2) + 4\left(1 - e^{-T_a/4}\right) \sin^2\left(\pi \Delta z^{(i)}/\lambda_s\right) \right], \quad (15)$$

that achieves maximum value at $\Delta z_{max}^{(i)} = \lambda_s/2$. In the considered case of the photon wavelength of 0.13 Å, corresponding to the energy 93.3 keV, $\Delta z_{max}^{(i)} = 6.5 \times 10^{-12}$ m.

In Fig. 2b, we plot the intensity (12) of the 93.3-keV photon stream at the exit from the resonant absorber displaced according to (13), (14). One of the following possible experimental realizations similar to [5,6] is considered. The source of 93.3-keV radiation emits a single spectral line of nearly natural width. The $^{67}$Zn absorber is a polycrystalline powder of ZnS, close to 100% enriched with $^{67}$Zn. It also has nearly natural single spectral line [4,7]. According to [7], the Lamb-Mossbauer factor at 4.2 K is $f_a \approx 1\%$, which provides the optical thickness $T_a = 1$ for the $^{67}$ZnS absorber length $L \approx 800$ μm. Taking into account the linear coefficient of non-resonant attenuation of 93.3-keV radiation in $^{67}$ZnS, $\mu \approx 1.88$ cm$^{-1}$ [25], one can estimate $T_e/T_a \approx 0.15$. As follows from (15), the maximum peak intensity of the produced pulses, achieved at $\Delta z^{(i)} = \lambda_s/2$

, depends on $T_a$ and $T_e$ and is maximized at $T_a \approx 3.2$ corresponding to the physical thickness of the $^{67}$ZnS powder $L \approx 2.56$ mm. Under these conditions, for any displacement of the source described by formulas (13) and (14), the approximation of fast motion is valid if the total time, $\Delta t_{total}^{(i)}$, meets the inequality $\Delta t_{total}^{(i)} \ll \tau_a \approx 2.9/(\gamma T_a) = 23$ µs. The longest reciprocating displacement of the source, shown in Fig. 2a, has a duration $\Delta t_{total}^{(6)} = 2\Delta t_f^{(6)} = 0.3$ µs. Thus, the approximate analytical solution (12) (the dotted green line in Fig. 2b) is very close to the result of numerical integration of equations (8)-(10) (the solid red line in Fig. 2b). As can be seen in Fig. 2b, the 93.3-keV radiation transmitted through the $^{67}$Zn absorber is similar to a digital sequence of optical pulses. It can be used, in particular, to implement a table-top nuclear quantum memory with 93.3-keV recoilless photons similar to the nuclear quantum memory with 14.4-keV photons implemented with a synchrotron source [26,27].

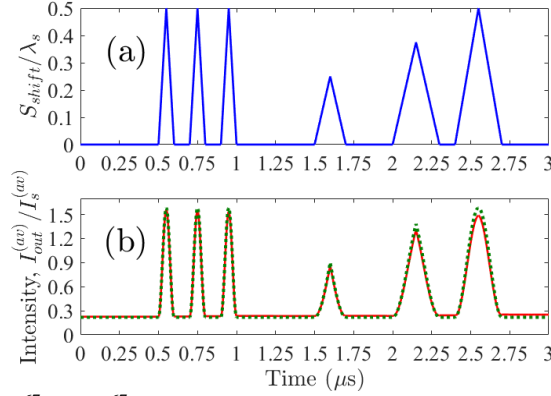

**Fig. 2.** Time dependence of the $^{67}$Ga or $^{67}$Cu source displacement function (13), (14) for $M = 6$ in panel (a) and the corresponding normalized intensity of the 93.3-keV photon stream at the exit from the resonant $^{67}$ZnS absorber in panel (b) with $T_a=3.2$ ($L = 2.56$ mm) and $T_e = 0.15T_a$. For the displacement function (13), (14), $t_{start}^{(1)} = 0.5$ µs, $t_{start}^{(2)} = 0.7$ µs, $t_{start}^{(3)} = 0.9$ µs, $t_{start}^{(4)} = 1.5$ µs, $t_{start}^{(5)} = 2$ µs, $t_{start}^{(6)} = 2.4$ µs, $\Delta t_b^{(i)} = \Delta t_f^{(i)} = 0.05$ µs ($i = 1,2,3$), $\Delta t_b^{(4)} = \Delta t_f^{(4)} = 0.1$ µs, $\Delta t_b^{(i)} = \Delta t_f^{(i)} = 0.15$ µs ($i = 5,6$), $\Delta t_c^{(i)} = 0$ ($i = 1,…,6$), $\Delta z^{(i)} = \lambda_s/2$ ($i = 1,2,3,6$), $\Delta z^{(4)} = \lambda_s/4$, $\Delta z^{(5)} = 3\lambda_s/8$. The green dotted and red solid lines are plotted using equations (12) and (8)-(10), respectively.

In Fig. 3 we also show the possibility of controlling the shape of individual pulses in a sequence. For example, it is possible to form a trapezoidal pulse (the first pulse in Fig. 3b), as well as asymmetrical triangular pulses (the second and third pulses in Fig. 3b).

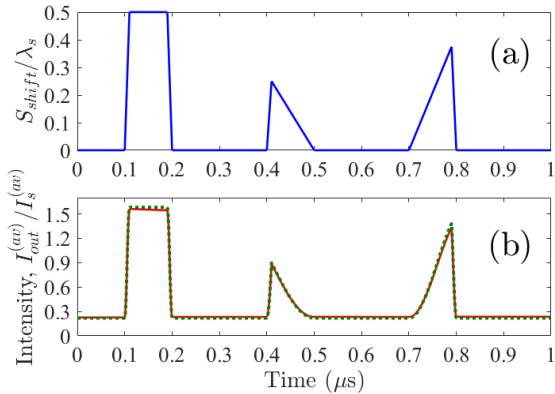

**Fig. 3.** Time dependence of the $^{67}$Ga or $^{67}$Cu source displacement function, $S_{shift}(t)$, (13), (14) for $M = 3$ in panel (a) and the corresponding normalized intensity of the 93.3-keV photon stream at the exit from the resonant $^{67}$ZnS absorber in panel (b) with $T_a=3.2$ ($L = 2.56$ mm) and $T_e = 0.15T_a$. For the displacement function (13), (14), $t_{start}^{(1)} = 0.1$ µs, $\Delta t_f^{(1)} = 0.01$ µs, $\Delta t_c^{(1)} = 0.08$ µs, $\Delta t_b^{(1)} = 0.01$ µs, $\Delta z^{(1)} = \lambda_s/2$, $t_{start}^{(2)} = 0.4$ µs, $\Delta t_f^{(2)} = 0.01$ µs, $\Delta t_c^{(2)} = 0$, $\Delta t_b^{(2)} = 0.09$ µs, $\Delta z^{(2)} = \lambda_s/4$, $t_{start}^{(2)} = 0.7$ µs, $\Delta t_f^{(3)} = 0.09$ µs, $\Delta t_c^{(3)} = 0$, $\Delta t_b^{(3)} = 0.01$ µs, $\Delta z^{(3)} = 3\lambda_s/8$. The green dotted and red solid lines are plotted using equations (12) and (8)-(10), respectively.

## V. CONCLUSION

In this paper, we have proposed a technique that makes it possible to transform radiation from $^{67}$Ga or modern $^{67}$Cu radioactive Mössbauer source with the photon energy of 93.3 keV of constant intensity into a sequence of short pulses with an arbitrary number of pulses, including a single pulse. The technique also allows one to individually and independently control, on demand, the moments of appearance, as well as the peak intensity, duration and shape of each pulse in the sequence. The technique is based on the transmission of Mössbauer (recoilless) photons of the source through a medium of resonantly absorbing nuclei $^{67}$Zn. The pulses appear due to rapid displacement of the source at some moments of time relative to the absorber, or vice versa, along the photon propagation direction by a distance less than the half-wavelength of the photon field, and return to its original position. When the source (or absorber) begins to move, the destructive interference between the incident field and the field coherently scattered by the nuclei, is transformed into constructive interference due to the Doppler effect. This leads to a sharp increase in intensity. When the source (or absorber) rapidly returns to its original position, destructive interference is restored, which causes a sharp decrease in intensity. As a result, a short pulse in the transmitted field intensity is appeared. The pulse begins with the beginning of the source (or absorber) displacement and finishes at the moment when the source (or absorber) stops in its original position. The peak intensity and shape of the pulse map the amplitude and time dependence of the source (or absorber) displacement. The pulse durations can be several orders of magnitude shorter than the lifetime of the emitting quantum transition of the source. As follows from [5,6,8-14,17,18], the γ-ray pulses with 93.3-keV photon energy of nanosecond duration can be produced with currently available equipment.

The proposed technique can expand the applications of Mössbauer spectroscopy and open up new prospects in X-ray quantum optics. It can, in particular, be used for implementation of a table-top nuclear quantum memory with 93.3-keV recoilless photons similar to the nuclear quantum memory implemented with a synchrotron source of 14.4-keV photons [26,27].

## ACKNOWLEDGMENTS

This work was supported by the Center of Excellence "Center of Photonics" funded by the Ministry of Science and Higher Education of the Russian Federation, contract 075-15-2020-906.